\documentclass[prd,
                reprint,
                letterpaper,
                groupedaddress,
                floatfix
            ]{revtex4-2}

\usepackage{graphicx}
\usepackage{color}
\usepackage{amsmath,amssymb}
\usepackage{acronym}
\usepackage[colorlinks=true]{hyperref}
\usepackage[dvipsnames]{xcolor}
\usepackage[normalem]{ulem}
\usepackage{booktabs}
\usepackage{blindtext}

\newcommand{\code}[1]{\texttt{#1}}

\renewcommand{\today}{\number\day\space\ifcase\month\or
  January\or February\or March\or April\or May\or June\or
  July\or August\or September\or October\or November\or December\fi
  \space\number\year}

\newcommand{\aref}[1]{\hyperref[#1]{the appendix}}

\newacro{SNR}{signal-to-noise ratio}

\newacro{BNS}{binary neutron star}

\newacro{BBH}{binary black hole}

\newcommand{\Mc}{\mathcal{M}}
\newcommand{\de}{\delta e}
\newcommand{\msun}{\ensuremath{M_{\odot}}}

\defcitealias{ly2017}{L\&Y 2017}

\begin{document}

\title{Method for detecting highly-eccentric binaries with a gravitational wave burst search}
\author{Belinda D.~Cheeseboro}
\email{bdc0001@mix.wvu.edu}

\affiliation{Center for Gravitational Waves and Cosmology, West Virginia University, Morgantown, WV 26505, USA}
\affiliation{Department of Physics and Astronomy, West Virginia University, Morgantown, WV 26505, USA}

\author{Paul T.~Baker}
\email[]{ptbaker@widener.edu}
\affiliation{Department of Physics and Astronomy, Widener University, Chester, PA 19013, USA}

\date{\today}

\begin{abstract}
Detection of gravitational waves (GW) from highly eccentric binary black hole (BBH) systems can provide insight into their dynamics and formation.
To date, all of the LIGO-Virgo BBH detections have been made using quasi-circular templates in their initial discovery.
However, recent studies have found some of these systems to be compatible with high eccentricity in the LIGO band, $e_{10 \textrm{Hz}} > 0.1$, possibly pointing to a population of sources that are challenging to detect.
Current low-latency search methods used with ground-based GW detector data are not well equipped to detect highly eccentric sources.
Template-based, matched-filter searches require accurate eccentric waveform models that are computational expensive.
Unmodeled burst searches are designed to detected localized excess power and are unable to identify multiple isolated bursts, as would originate from a single highly eccentric BBH.
Therefore, we propose a signal-based prior that can be incorporated into an existing GW burst search to target highly eccentric BBHs.
Our eccentric burst prior is based on the Newtonian burst model described by \cite{ly2017}.
As a proof of concept, we test our method on simulated data and find that for intermediate SNR $\sim3-6$ signals using the eccentric burst prior more effectively localizes GW bursts when compared to a uniform prior.
\end{abstract}


\maketitle


\section{Introduction}
\label{sec:intro}
In the past several years, a plethora of gravitational wave (GW) events have been detected by the Advanced LIGO \cite{TheLIGOScientific:2014jea} and Virgo \cite{TheVirgo:2014hva} observatories \cite{LIGOScientific:2018mvr,LIGOScientific:2020ibl,LIGOScientific:2021usb}. In particular the detections of GW190521, GW200105, and GW200115 have improved our understanding of the stellar graveyard population by pushing the limits of previous population models \cite{LIGOScientific:2020iuh,LIGOScientific:2021qlt}.
To date, all of the GW events listed in LIGO-Virgo's Gravitational Wave Transient Catalogue have been initially identified using quasi-circular waveform templates \cite{LIGOScientific:2018mvr,LIGOScientific:2020ibl,LIGOScientific:2021usb}.

The recovered eccentricity of a GW source is one of several source parameters that can provide insight into the environment where the binary formed.
Environments in which binary black holes (BBH) form can be broken into two categories: isolated and dynamic \cite{Mapelli:2021taw}.
BBHs that formed in an isolated environment steadily decay in eccentricity, circularizing over time \cite{Peters:1964zz}.
This means that BBHs formed in isolation will most likely enter the LIGO-Virgo band with little to no eccentricity.
BBHs that formed in a dynamic environment, such as globular clusters, are more likely to enter the LIGO-Virgo band with detectable eccentricity ($e_{10 \textrm{Hz}} > 0.1$, where $e_{10 \mathrm{Hz}}$ is the binary eccentricity at a reference frequency of $10$ Hz) \cite{Samsing:2017rat,Samsing:2013kua, Rodriguez:2017pec,Rodriguez:2018pss,Zevin:2021rtf}.

Further studies of LIGO-Virgo detections covering the first and second observing runs show that the GW sources are consistent with little to no eccentricity, $e_{10 \textrm{Hz}} < 0.1$ \cite{Romero-Shaw:2019itr,Wu:2020zwr,OShea:2021ugg}. 
However, measurements of the eccentricity of GW190521 and GW190620A strongly suggest it contains $e_{10 \textrm{Hz}} > 0.1$  \cite{Gayathri:2020coq,Romero-Shaw:2020thy,Romero-Shaw:2021ual}, providing a strong argument for the inclusion of eccentricity in current low-latency searches.

One of the main methods that LIGO-Virgo-Kagra Collaboration uses for detection runs on GW data is matched-filtering \cite{2012ApJ...748..136C,2016CQGra..33u5004U,2012PhRvD..85l2006A}. Matched-filtering is a highly sensitive method that requires accurate waveform models to generate templates to use for analysis.
It is expected that a large fraction of the eccentric source population will enter the LIGO-Virgo band with $e_{10 \mathrm{Hz}} \lesssim 0.1$.
Even within this low-eccentricity population, there is a small fraction ($\sim 5\%$) of sources that are not detectable using quasi-circular waveform models.
For sources with $e_{10 \mathrm{Hz}} > 0.1$, eccentric waveform models are needed to detect the complete population \cite{Zevin:2021rtf}.
Even when eccentric sources are detectable using quasi-circular waveforms, the mismodeling can bias parameter estimation \cite{OShea:2021ugg}.

Currently, there are a variety of waveforms that use eccentricity in their models \cite{Moore:2019xkm,Moore:2016qxz,Cao:2017ndf,Huerta:2016rwp,Tiwari:2019jtz,Tanay:2016zog,Hinder:2017sxy,Huerta:2014eca,Huerta:2017kez,Nagar:2021gss}.
Of the available eccentric waveforms only a few that have been implemented into \code{LALsuite} \cite{lalsuite} model only the inspiral phase of the signal, lacking the merger and ringdown \cite[e.g.][]{Huerta:2014eca,Moore:2016qxz,Tanay:2016zog}.
Not having a complete waveform template can exclude information that is important for constraining the source parameters of BBH systems.
The dearth of eccentric waveforms make it difficult to conduct template-based searches for eccentric BBHs, especially highly eccentric BBHs.

In addition to  matched-filter searches LIGO/Virgo also employs unmodeled burst search methods, such as \code{BayesWave} \cite{cl2015} and Coherent Wave Burst (cWB) \cite{kymm2008}.
As there are a lack of reliable eccentric waveforms implemented into \code{LALsuite}, LIGO/Virgo used cWB to search for eccentric BBH in O1 and O2 data. 
With cWB they recovered 7 out of the 10 BBH detections, which they believe to be consistent with little to no eccentricity, as in the template-based search.
No new eccentric BBH candidates were identified \cite{Salemi:2019owp}.
These burst searches are not as sensitive as a template-based search, but they are not constrained by having to use waveform models in their analyses.
Currently, these burst search methods are designed to look for excess power in compact regions of time-frequency space. 
This makes them a bad fit for detecting the temporally disconnected bursts from highly eccentric BBH.
Given a marginal signal-to-noise ratio GW from a highly eccentric BBH, a traditional burst search like \code{BayesWave} is likely to detect the loud merger, but miss the quieter bursts emitted during the inspiral phase \cite{2017APS..APR.T1012C}.
With the merger alone it may be impossible to determine that the system had significant eccentricity earlier in its evolution.

We propose a method that could be implemented within the existing GW burst pipelines to search for the temporally disconnected bursts from highly eccentric BBH.
This method is largely based on a model described by Loutrel \& Yunes \cite[][hereafter \citetalias{ly2017}]{ly2017}.
In \autoref{sec:model} we extend the GW burst evolution from \citetalias{ly2017} to include both forward and backward evolution in time.
We use the evolution equations to define an eccentric burst prior, and we describe a simplified signal model to use with the prior.
Next, in \autoref{sec:recov} we described two sets of simulations that were use to test our methods and the results of those tests.
Finally, in \autoref{sec:discuss} we summarize our findings, including the limitations of this method and modifications that could improve it.

\section{Modeling Bursts from Highly Eccentric Sources}
\label{sec:model}
Our method is based on the model described by \citetalias{ly2017}.
They model the GWs emitted by highly eccentric binaries as a series of discrete GW bursts.
The bursts are assumed to be emitted instantaneously at each pericenter crossing of the system with negligible emission occurring during the rest of the orbit.
In the Newtonian burst model the system follows a series of Keplerian orbits.
The orbital parameters are constant throughout each orbit, updating after the burst of GWs is emitted.
The instantaneous eccentricity refers to the eccentricity of the current orbit used to determine the GW emission at a particular pericenter crossing.

\par
Each GW burst is assumed to be compact in time and frequency.
\citetalias{ly2017} represent each burst as a rectangular tile in time-frequency space, characterized by its centroid, $t,f$, and two widths, $\sigma_t, \sigma_f$.
They evolve the binary system forward in time using a slow speed post-Newtonian expansion.
Given the time-frequency location of an initial burst of GWs, their model uses the total mass of the binary, $M$, pericenter separation, $r$, and instantaneous eccentricity, $e$, to determine the time until the next burst, $\Delta t$, and update the orbital parameters $r$ and $e$ for the next orbit.
These orbital parameters also determine the gravitational wave frequency of each burst.
\par
As a proof of concept, our method is based on the lowest order evolution presented by \citetalias{ly2017}: Keplerian orbits perturbed by the instantaneous emission of quadrupolar gravitational radiation at pericenter.
Instead of using rectangular time-frequency tiles, we represent each burst as a single Morlet-Gabor wavelet, imitating \code{BayesWave} \cite{cl2015}.
We use the evolution model of \citetalias{ly2017} to place a joint prior on the centroids of all wavelets.
Model wavelets that are located in time-frequency space according to the \citetalias{ly2017} evolution are assigned high prior probability, while low prior probability is assigned to wavelets elsewhere (e.g. between two pericenter crossings).

This results in a signal description that is not a fully phase-connected waveform model.
The model we present has more freedom to overcome systematic waveform uncertainties at the expense of some lost signal power.

\subsection{Centroid Mapping Equations}
\label{subsec:cm_eq}
\citetalias{ly2017} defines the orbital parameters of one burst in terms of the burst that directly precede it.
Their expansion relies on two small parameters associated with pericenter separation $\frac{M}{r}\ll 1$ and eccentricity $\de = 1-e \ll 1$.
For the remainder of this section we will work in geometrized total mass units, where $G=c=1$ and the binary total mass $M=1$.
Chirp mass $\Mc$, distance $r$, and time $t$ are measured in units of $M$ and frequency in units of $M^{-1}$.

The forward orbital evolution of a highly eccentric binary system is presented in section 2.1 of \citetalias{ly2017}, to leading order.
This gives the orbital parameters associated with the next ($i+1$) GW burst in terms of the previous ($i$).
In addition to the forward evolution, we want to know the backwards evolution of the system.
If we detect one burst from a highly eccentric binary, we want to search for other GW bursts produced by the same system.
To obtain the backwards evolution we can invert their solution, solving for the previous (${i-1}$) burst in terms of the next $(i)$.
Working to first order

\begin{subequations} \label{eq:re.evo}
\begin{align}
    \de_{i\pm 1}  &\approx \de_{i} \pm \frac{85 \pi \sqrt{2}}{12} \Mc^{5/3} r_{i}^{-5/2} \left(1 - \frac{1718}{1800} \de_{i}\right) \label{eq:de.evo} \\
    r_{i\pm 1} &\approx r_{i} \left[1\mp\frac{59 \pi \sqrt{2}}{24} \Mc^{5/3} r_{i}^{-5/2} \left (1+\frac{121}{236}\de_{i} \right) \right]. \label{eq:r.evo}
\end{align}
\end{subequations}

The orbital evolution of the binary can be mapped to the gravitational wave signal evolution in time-frequency space. In the Newtonian burst model the time between bursts is the time between pericenter crossings, i.e.~the orbital period.
The burst frequency is set by the instantaneous orbital frequency defined by the pericenter distance and orbital speed $2\pi f_{GW}\sim \frac{v}{r}$.
The period and pericenter speed are given by Kepler's laws to leading order, so
\begin{subequations}\label{eq:orb2GW}
\begin{align}
    t_{i} &\approx t_{i-1} + 2\pi\sqrt{\frac{r_i^3}{\de_i^3}} \\
    2\pi\, f_i &\approx \sqrt{\frac{(2-\de_i)}{r_i^3}}.
\end{align}
\end{subequations}
Recall that $M=1$ in our total mass units.
In both relations of \autoref{eq:orb2GW} the orbital parameters of the $i$th burst on the right-hand-side are determined from the previous burst using \autoref{eq:re.evo}.

Given the gravitational wave parameters of a burst, we first determine the orbital parameters.
We then evolve the orbital parameters to the next (or previous burst) and
finally determine the gravitational wave parameters of that burst.

For our analysis, any burst can be fully characterized by three parameters: its central time, central frequency, and instantaneous eccentricity.
(The pericenter distance makes a convenient computational midpoint, but we prefer to work directly with gravitational wave frequency in the model.)
If these three parameters are known for any one burst, the binary's chirp mass will uniquely determine all other bursts where our leading order approximations are valid.
This allows us to directly write the leading order evolution in terms of only $t$, $f$, and $\de$.
Given the $i$th burst the next burst in the sequence is
\begin{widetext}
\begin{subequations} \label{eq:tffor}
\begin{align}
    t_{i+1} &= t_{i} + \frac{1}{f_{i}} \sqrt{\frac{2-\de_{i}}{\de_{i}^3}}
        \label{eq:tfor} \\
    f_{i+1} &= f_{i} \left[ 1+\frac{23}{2} \left( \frac{\pi}{2} \right)^{2/3} (\pi f_{i} \Mc)^{5/3} \left(1+\frac{7547}{4140} \de_{i} +\frac{3725}{552} \left( \frac{\pi}{2} \right)^{2/3} (\pi f_{i} \Mc)^{5/3} \right) \right] \\
    \de_{i+1} &= \de_{i} + \frac{85}{3} \left( \frac{\pi}{2} \right)^{2/3} (\pi f_{i} \Mc)^{5/3} \left(1 - \frac{121}{225} \de_{i} \right),
\end{align}
\end{subequations}
and the previous burst in the sequence is
\begin{subequations}\label{eq:tfback}
\begin{align}
    t_{i-1} &= t_{i} -\frac{1}{f_{i}} \sqrt{\frac{2-\de_{i}}{\de_{i}^3}}
        \label{eq:tback} \\
    f_{i-1} &= f_{i} \left[ 1-\frac{23}{2} \left( \frac{\pi}{2} \right)^{2/3} (\pi f_{i} \Mc)^{5/3} \left(1+\frac{7547}{4140} \de_{i} +\frac{5875}{368} \left( \frac{\pi}{2} \right)^{2/3} (\pi f_{i} \Mc)^{5/3} \right) \right] \\
    \de_{i-1} &= \de_{i} - \frac{85}{3} \left( \frac{\pi}{2} \right)^{2/3} (\pi f_{i} \Mc)^{5/3} \left(1 - \frac{121}{225} \de_{i} - \frac{295}{12} \left( \frac{\pi}{2} \right)^{2/3} (\pi f_{i} \Mc)^{5/3} \right).
\end{align}
\end{subequations}
\end{widetext}
Ideally, the forward and backward evolution would be perfect inverses of each other.
However, because we truncate the evolution to low order, this is not the case.
Owing to the low order approximation, we find the error in the time-frequency evolution to be less than $10\%$ for $e \lesssim 0.7$ and $r \gtrsim 15M$.

\subsection{Eccentric Burst Prior}
\label{subsec:prior}
The centroid mapping procedure tells us where bursts are expected to occur.
We wish to construct a prior for the observed burst centroids given their expected locations.
The eccentric burst prior consists of a sum of bivariate normal distrubutions, each centered on an expected burst centroid.
The covariance of each peak, $\Sigma_i$, represents the uncertainty in the wavelet being centered at that location.

The prior is defined by a set of five meta-parameters: the binary total mass, $M$, and chirp mass, $\Mc$, the gravitational wave signal parameters of one burst, $(t_\star, f_\star)$, referred to as the ``anchor burst'', and the instantaneous eccentricity of the system for the orbit that produced the anchor burst $\de_\star$.
\begin{subequations} \label{eq:prior}
\begin{align}
    p(t,f;\, M,\Mc,t_\star,f_\star,\de_\star) &= \frac{1}{2\pi\,N} \sum_{i=0}^N \frac{1}{\det\Sigma_i} e^{-\mathbf{d}_i \Sigma_i^{-1} \mathbf{d}_i^T} \\
    \mathbf{d}_i &= [t, f] - [t_i, f_i],
\end{align}
\end{subequations}
where $N$ is the number of bursts expected in the observation time to be analysed.
The total mass does not appear in the centroid mapping \autoref{eq:re.evo}.
It is effectively a unit conversion factor, converting the burst $t,f$ from the S.I. units of our analysis to the total mass units of the centroid mapping equations.

Starting from the covariance of the anchor burst peak, $\Sigma_\star$, the uncertainty can be propagated through the centroid mapping equations using a Jacobian.
This allows us to determine the covariance of all other peaks in the prior distribution.
To account for the dependence of $t$ and $f$ on $\de$, we compute the covariance evolution in three dimensions, but only use the appropriate 2D block when constructing the prior.
\begin{subequations} \label{eq:sigma}
\begin{align}
    \mathbf{x} = (t, f, \de) \\
    J_{\pm, mn} = \frac{\partial x_{m,i}(\mathbf{x}_{i\pm1})}{\partial x_n} \\
    \Sigma_{i} = J_{\pm} \Sigma_{i\pm 1} J^{T}_{\pm},
\end{align}
\end{subequations}
where $i$ is the index of the peak (i.e., which burst) and the $m,n$ indices of the Jacobian matrix run through the three burst parameters.
$x_{m,i}$ are the parameter evolution equations, \autoref{eq:tffor} and \autoref{eq:tfback}, which are functions of the $(i\pm 1)$th parameters.
The full functional forms of $J_\pm$ are presented in \aref{app:jacobian}.

\begin{figure}[t]
    \includegraphics[width=\columnwidth]{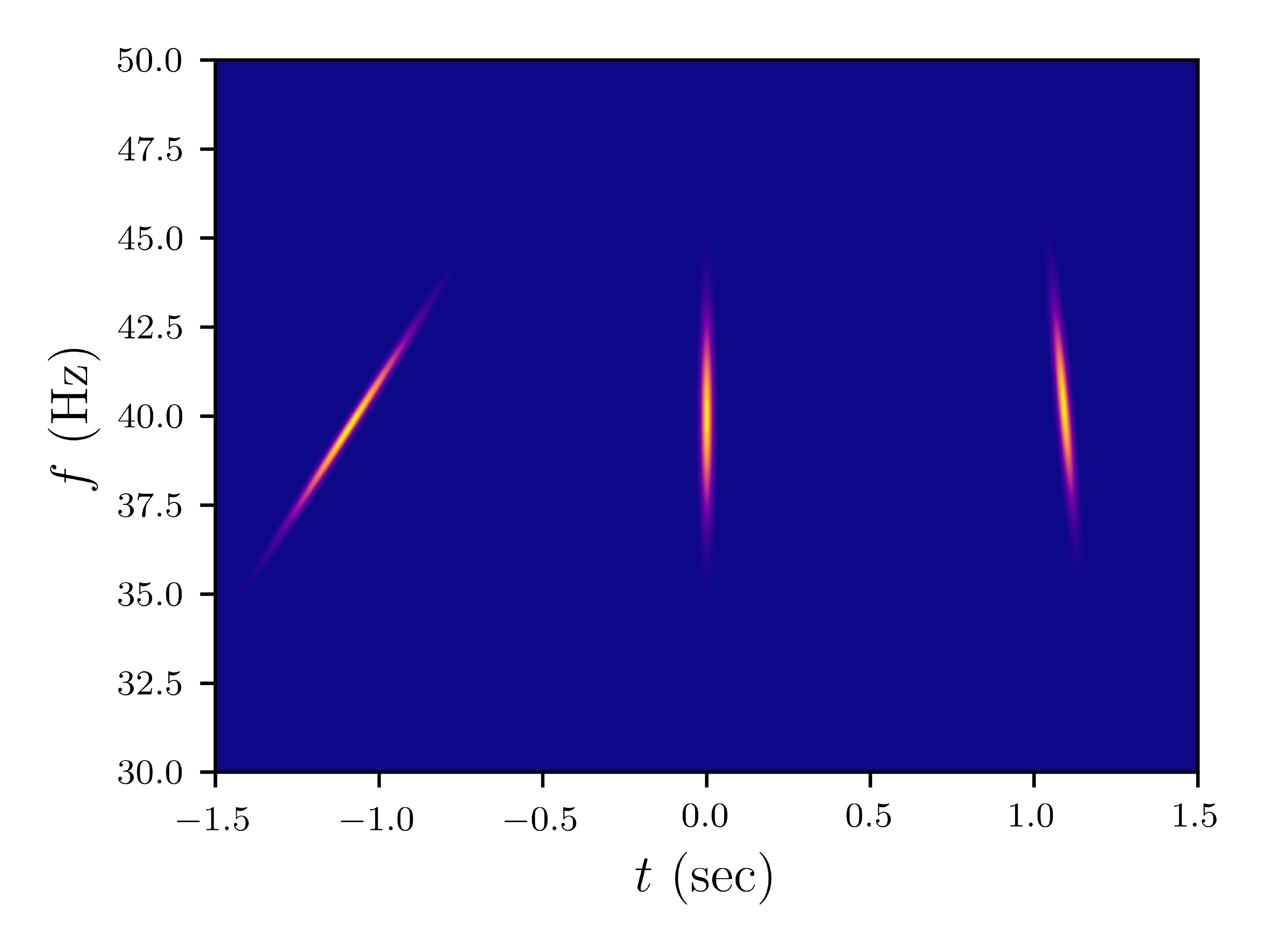}
    \caption[Prior probability as a function of time-frequency location.]{Prior probability for the time-frequency location ($t,f$) of the centroid of a gravitational wave burst for the meta-parameters $M=50\msun$, $\Mc=15\msun$, $t_\star=0$, $f_\star=40$ Hz. $\de_\star=0.1$.
    In this case the prior preference is to have three bursts in the analyzed time window.
    The full prior is the sum of three bivariate normal peaks, one for each expected burst.
    \label{fig:prior_ex}}
\end{figure}

A 2D time-frequency map of the prior is shown in \autoref{fig:prior_ex} for a choice of meta-parameters.
For these parameters, three bursts are expected occur during the $3$ s analysis window.
More bursts should occur both before and after these three, but they do not contribute to the prior.
The central peak is the location of the anchor burst, and is vertically aligned.
The Jabobian evolution causes the $t$, $f$, and $\de$ uncertainties to mix introducing $t$-$f$ covariance in the other peaks.
Intuitively, the tilting of the outer peaks toward the center is primarily driven by the uncertainty in $f$.
High frequency sources will have less time between bursts and low frequency sources will have more time.
The time between the high frequency edges of the peaks is reduced, relative to the centroid and the time between the low frequency edges is increased.

In practice $(t_\star, f_\star)$ should closely follow the $(t,f)$ of one detected burst.
There is some ambiguity in which is the anchor burst.
Assuming the centroid mapping is approximately symmetric, any burst in the series could act as the anchor burst with relatively even probability.
This leads to the meta-parameter distributions being multimodal and presents a practical challenge when sampling the space.

\subsection{Signal Model}
\label{subsec:sig_model}
To accompany the eccentric burst prior, we require a GW burst signal model.
We implement a simplified version of the \code{BayesWave} model.
\code{BayesWave} models excess detector power using Morlet-Gabor wavelets \cite{cl2015}.
Our signal model assumes the GWs emitted at each pericenter crossing can be fit with a single wavelet.
Each wavelet is defined by five parameters,
\begin{multline}
    \Psi_i (t; t_i, f_i, A_i, Q_i, \phi_i) = A_i\, \exp\left[-\left(\frac{2\pi f_i}{Q_i}\right)^2(t-t_i)^2 \right] \\
    \times \cos{\left[2 \pi (t-t_i) + \phi_i\right]},
    \label{eq:wavelet}
\end{multline}
where $A_i$ is its amplitude, $Q_i$ its quality factor, $\phi_i$ its reference phase, and $t_i$ and $f_i$ are the central time and frequency of the wavelet.
The subscript $i$ refers to the individual wavelets in the model, where the full signal model is the sum of all wavelets.
\par

The eccentric burst prior is a prior on the wavelets' time-frequency centroids, $t_i$ and $f_i$.
A given signal model will have a number of $(t,f)$ pairs equal to the number of wavelets in the model.
The expectation is that each high probability peak in the eccentric burst prior should have one wavelet nearby.
The eccentric burst prior associates multiple temporally disconnected bursts into a single GW signal with one wavelet per burst.
This goes against the usual GW burst search techniques which focus on reconstructing a single GW burst with multiple wavelets that are nearby in time-frequency space.
\par

The GW bursts emitted during pericenter crossings of a highly eccentric orbit contain only one GW cycle.
We expect the recovered wavelets to have low quality factors $Q\sim 2$, although we do not formally enforce this expectation with a prior on $Q$.
\code{BayesWave} uses a Reversible Jump Markov Chain Monte Carlo (RJMCMC) to sample not only the individual wavelet parameters but also the number of wavelets in the model.
Implementing a comparible RJMCMC scheme to \code{BayesWave} is beyond the scope of this proof of concept study. For simplicity, we opt to fix the number of wavelets in our signal model.

\section{Analysis of Simulated Data}
\label{sec:recov}
To test our eccentric burst prior method, we simulate single detector LIGO/Virgo-like data.
This simulated data does not incorporate the response of any particular detector, assuming an optimally oriented source and a white noise spectrum across the whole range of frequencies simulated.

The injected GW signals cover a narrow frequency band, so the lack of a realistic noise spectrum is not a major issue.
We use a Gaussian, white noise likelihood.
While the eccentric burst prior is not as prescriptive as a phase-connected waveform, it still works best when the small parameter conditions of the centroid mapping series are met.
These conditions are satisfied for GW bursts emitted at earlier times of the inspiral.
The simulated signals are truncated to contain three GW bursts emitted well enough before merger that these conditions are met.
In its current form our model would have trouble with bursts emitted closer to merger.  The approximations used to determine \autoref{eq:tffor} and \autoref{eq:tfback} would not be valid.  Additionally, the wavelets would preferentially fit the higher amplitude merger, missing the lower amplitude early inspiral.

Since each simulation contains three bursts, we fix the number of wavelets in the \code{BayesWave}-like signal model to three.

We perform two sets of simulations using both a uniform prior and the eccentric burst prior.
The first uses simple waveforms constructed from three wavelets spaced according to the centroid mapping equations of \autoref{subsec:cm_eq}.
These simple waveforms contain bursts that are roughly 20 orbits before merger.
The second uses a more realistic waveform model for highly eccentric BBHs \cite{2013PhRvD..87d3004E}.
The bursts in this set are roughly 80 orbits before merger.
In each case we use sources with an initial eccentricity of $e=0.9$, $\de=0.1$ and cover GW frequencies of $\sim30-65$ Hz.
\par

For both the eccentric burst prior and uniform prior cases, we place uniform priors on the wavelets' quality factors ($1.5 \le Q \le 15$), phases ($-\pi \le \phi \le \pi$), and amplitudes ($A>0$).
The uniform priors for the wavelets' central time and frequency cover the entire time-frequency data volume, as set by the time span and sampling rate of the data.
Each simulation contains $1.5 \lesssim T \lesssim 3$ s of data.
The corresponding frequency prior is uniform on $1/T \le f \le f_\mathrm{Nyquist}$.
The eccentric burst prior is set to zero outside the same time-frequency volume as the uniform prior.
\par

We place uniform priors on the meta-parameters of the eccentric burst prior.
Like the wavelet time and frequency parameters, the priors for anchor burst time and frequency are uniform on the whole data volume.  The anchor burst eccentricity is uniform on $10^{-3} \le\de\le 0.9$.
The prior for total mass is uniform on $ 0 \le M \le 200\msun$.
The chirp mass is held to $\Mc>0$.
Because $\Mc< M$, the prior on $M$ additionally constrains $\Mc$.

To evaluate the efficacy of the eccentric burst prior, we use a parameter estimation figure of merit (\autoref{fig:toy_east_uncert}).
To better compare between individual simulations, we use the fractional uncertainty in recovered time $\Delta t \times f_\mathrm{inj}$, where $\Delta t$ is the standard deviation of the marginal posterior distribution for the wavelet central time and $f_\mathrm{inj}$ is the injected GW frequency.

In the low signal-to-noise ratio (SNR) limit any GW signal should be undetectable, regardless of whether the eccentric burst prior or a uniform prior is used.
In these cases the recovered $t$ should be unconstrained with its posterior filling the entirety of the prior range.
The time uncertainty $\Delta t$ should be of the order of the time span of the data being analyzed, which defines the full prior range.

In the high SNR limit, a GW signal should be easily detectable, and the posterior probability distribution will be dominated by the likelihood (assuming the prior includes the maximum-likelihood probability region of the parameter space).
In these cases regardless of whether one uses the eccentric burst prior or a uniform prior, the parameter estimation uncertainty should be the same, as it just the spread of the marginal posterior distribution.
Between these two limits lies a transition region, where the choice of prior affects the parameter estimation uncertainty.

We choose to use the parameter estimation uncertainty as a proxy for a true detection statistic.
Since the two priors, eccentric burst and uniform, have the same limiting behavior, we can examine at what SNR each prior converges to these limits.

We opt for this simplified detection proxy because common information criterion statistics are unsuitable for our problem.
For example the Akaike information criterion (AIC) uses the maximum likelihood and does not take into account priors.
The deviance information criterion (DIC) is constructed from the variance of the posterior samples and is ill-defined for cases where with mode switching \cite{gcs+2014}.
We observe mode switching in the intermediate SNR regime of interest, for example where the sampler switches back and forth from detecting two of the injected bursts to detecting all three.

The other model parameters are not as well suited to the task of tracking detection.
This is because of covariances between the parameters.
In addition to time uncertainty, we present the fractional uncertainty in recovered GW frequency $\Delta f / f_\mathrm{inj}$.
This figure of merit has the same high-SNR limiting behaviour as $\Delta t\times f_\mathrm{inj}$, where the choice of prior does not affect the uncertainty.
At low SNR $f$ is not fully unconstrained by the eccentric burst prior because of the covariance between $f$ and $M$.
The binary total mass effectively sets the merger frequency, so $f$ is held to lower values for high mass systems. (This effect can be seen in \autoref{fig:corner} in the 2D marginal posterior for anchor burst frequency, $f_\star$, and total mass, $M$.)

\subsection{Simple Wavelet Injections}
\label{subsec:toy}
The first set of simulations use injected waveforms based on the centroid mapping equations found in \autoref{subsec:cm_eq}.
Starting from a randomly selected total mass in the range {$M=30-70\msun$} and fixed mass ratio $q=0.3$ (which in turn determine the chirp mass), we use \autoref{eq:tffor} to determine the time-frequency locations of three GW bursts.
We inject a $Q=2$ wavelet at each of these locations.
We simulate {13} sources covering a frequency range of $f = 30-60$ Hz.
For each source we scale the amplitude to achieve a target SNR, allowing us to search for the same source at a range of SNRs from $2$ to $7$.

In order to keep the expected number of bursts in our analysis window at exactly three, we truncate the data before the next burst was expected.
This results in unequal observation times ranging from about \hbox{$2-3$ s}.

\begin{figure*}[t]
    \includegraphics[width=\textwidth]{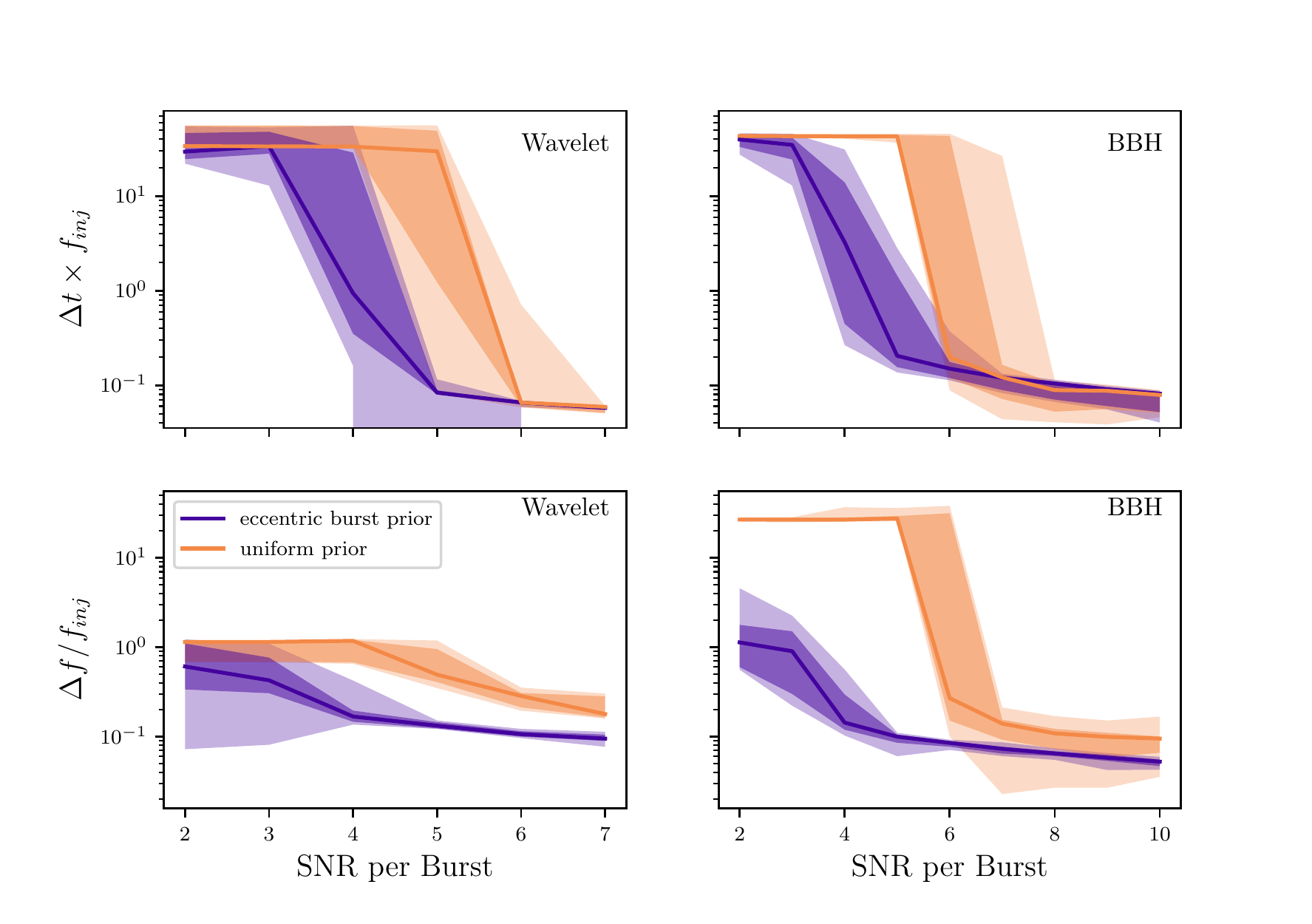}
    \caption{Parameter estimation uncertainty for the simple wavelet (\textit{left}) and eccentric BBH waveform (\textit{right}) injections as described in \autoref{subsec:toy} and \autoref{subsec:wave}. The solid lines show the median uncertainty of all simulations, and the dark and light shaded regions represent the 68\% and 95\% credible intervals on the uncertainty, respectively. At low SNR the posterior distributions should fill the prior range, leading to large uncertainty. At high SNR the posterior distributions are localized leading to small uncertainty. Using the eccentric burst prior leads to better signal recovery in the intermediate SNR regime $\sim 3-6$.}
    \label{fig:toy_east_uncert}
\end{figure*}

We analyze each simulation twice, first with the eccentric burst prior and again using a uniform prior on the individual wavelets' time-frequency locations.
We use \code{PTMCMCSamper} to draw posterior samples  \cite{justin_ellis_2017_1037579}.
From these we determine our two parameter estimation figures of merit for each simulation.
These figures of merit are summarized in the left panels of \autoref{fig:toy_east_uncert}, combining all trials.

As we expect, $\Delta t\times f_\mathrm{inj}$ converges to the same low and high SNR limits regardless of which prior is used.
For $\mathrm{SNR} \le 3$, the injected signals are not detected.
The posterior distribution for each wavelet's central time is approximately uniform across the entire observation, leading to large uncertainties in all trials.
Using the eccentric burst prior, the injected signals are reliably detected at $\mathrm{SNR} \approx 5$, a lower amplitude than with the uniform prior.
At the highest SNRs the posterior is dominated by the likelihood, so all bursts are detected with effectively the same fractional uncertainty.

The frequency uncertainty, seen in the lower left panel of \autoref{fig:toy_east_uncert}, is less clean.
At higher SNRs the eccentric burst prior is more constraining than the uniform prior, meaning the posterior is not fully likelihood dominated.

Using the eccentric burst prior results in improved parameter estimation precision in the intermediate SNR regime (SNR $\sim 3-6$), as can be seen in the upper left panel of \autoref{fig:toy_east_uncert}.
This improved detection confidence is especially important when trying to detect quiet bursts from highly eccentric BBHs during the early inspiral phase.
These initial results set a benchmark for further tests using a more realistic GW waveform.

\subsection{Eccentric BBH Waveform Injections}
\label{subsec:wave}
We follow the same basic framework as \autoref{subsec:toy} to work with more realistic eccentric BBH waveforms from \cite{2013PhRvD..87d3004E}.
This model maps the binary dynamics to an effective single body following a Kerr geodesic.
It was designed to approximate GWs from highly eccentric systems that resulted from the radiation-induced dynamical capture of initially hyperbolic orbits \cite{2013PhRvD..87d3004E}.
As in the simple wavelet model, we simulate multiple waveforms with initial eccentricity of $e=0.9$, $\de=0.1$.
We choose an initial pericenter separation of $r=15M$ to stay within the range of validity of the low order centroid mapping equations used in the eccentric burst prior.
Unlike in the wavelet injections, we select {16} total masses in the range $M=15-30\msun$, resulting in waveforms with GW frequencies of $f=30-65$ Hz.
Again we truncate the waveforms to contain exactly three bursts from the early inspiral, giving observation times of \hbox{$1.5-3$ s}.
These waveforms use a larger sampling rate by a factor of $\sim10$ resulting in a significantly wider analysis bandwidth compared to the previous simulations.

We assume a detector orientation aligned with the plus polarization of the GWs.
The waveform amplitudes were directly scaled to set the injected SNR.
The waveform amplitudes were scaled to control the injected SNR by first calculating the baseline SNR of the three bursts within our time window and then taking the fraction of the desired injected SNR with the baseline SNR to create a factor to appropriately scale the signal.
This is effectively equivalent to changing the distance to the source but does not account for cosmological redshift of the system mass.

These more realistic waveforms have more structure than the wavelets used in the previous simulations.
Fitting each burst with a single wavelet results in an imperfect recovery.
This leads to the recovered SNR being less than the injected, so we use an extended range of SNRs compared to the simple wavelet injections, going up to $10$.

As before we analyze each simulation twice, collecting posterior samples first with the eccentric burst prior and again without.
The same parameter estimation figures of merit are plotted as a function of SNR in \autoref{fig:toy_east_uncert}.
Looking at the top right panel of \autoref{fig:toy_east_uncert}, we observe the same mode switch behavior as in top left panel \autoref{fig:toy_east_uncert} where the sampler switches from non-detection to detection in the intermediate SNR range $3-6$.
From looking at this intermediate region (SNR range $3-6$), we see a similar improvement in the fractional uncertainty of recovered burst central times when the eccentric burst prior is used compared to the uniform prior.

The constraining effect of the $M$-$f$ covariance is more apparent in this case when looking at the bottom right panel of \autoref{fig:toy_east_uncert}.
(The $M$-$f$ covariance can be clearly seen in the bottom-leftmost plot in \autoref{fig:corner}.)
The higher Nyquist frequency of the BBH waveform injections inflated the recovered frequency uncertainty in the low SNR, uniform prior case relative to the wavelet injections.
At low SNR the eccentric burst prior constrains the signal frequency to stay in the low $f$ part of the search parameter space.

\section{Discussion}
\label{sec:discuss}
We develop an eccentric burst prior that can be used to improve the detection prospects for highly eccentric BBH systems when used with a gravitational wave burst search.
In particular the eccentric burst prior improves performance for intermediate SNR bursts generated during the inspiral phase of the BBH evolution, which may be missed by traditional burst methods that focus on the merger.
As a proof of concept, we verify this improvement using simulated LIGO/Virgo-like data.  From these simulations we found that the largest improvement over a uniform prior occurred for burst SNRs in the $3-6$ range.

\begin{figure*}[tp]
    \includegraphics[width=\textwidth]{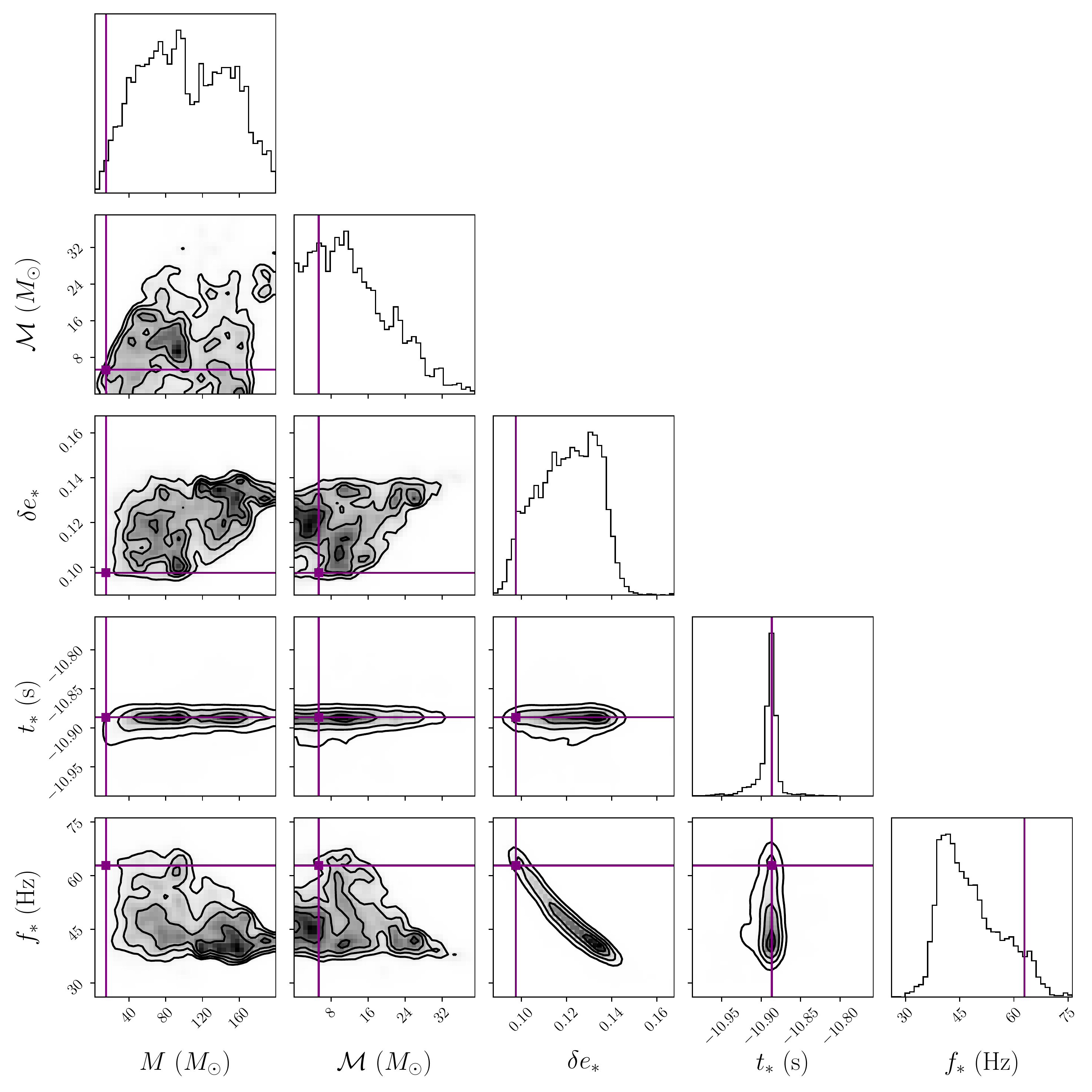}
    \caption{Corner plot showing the posterior distributions for the eccentric burst prior meta parameters for a case with $\textrm{SNR per Burst = 4}$.
    This simulation used the highly eccentric waveform model of \cite{2013PhRvD..87d3004E} with signal parameters $M=15\msun$ and $\Mc=5.3\msun$.
    The anchor burst parameters $(t_\star, f_\star, \de_\star)$ are related to the system orbital parameters at the emission of the anchor burst.
    The time, $t_\star$, is measured relative to merger.
    \label{fig:corner}
    }
\end{figure*}

The eccentric burst prior is based on a model for the time-frequency location of bursts that depend on the physical parameters of the BBH system.
The total mass $M$ and chirp mass $\Mc$ act as meta-parameters of the prior.
We hope that the posteriors for these parameters would carry useful information about the GW system producing the observed GWs.
Unfortunately the recovery of these parameters was very poor, as can be seen in \autoref{fig:corner}.

Since we analyzed only three bursts from the earlier part of the inspiral of each simulation, we have limited the information to work with.
The GW frequency does not evolve significantly over these three bursts, meaning we could not constrain the chirp mass $\Mc$, which relates to the frequency derivative as $\mathcal{M} \sim \dot{f}^{3/5}$.

The total mass $M$ would be better constrained by measuring the merger frequency, which we do not have access to.
In addition to the system masses the anchor burst parameters $(t_\star, f_\star, \de_\star)$ give additional degrees of freedom.
This means there are many combinations of meta-parameters that lead to appropriately placed bursts.

If we analyze the whole waveform all the way up to the merger, we would expect more meaningful meta-parameter recovery, but that presents additional challenges.
The centroid mapping equations, \autoref{eq:tffor} and \autoref{eq:tfback}, were based on the lowest order orbital evolution presented in \citetalias{ly2017}.
This low order approximation breaks down well before the merger.
If we try to include bursts near the merger, the prior disagrees with the reality.
The peaks in the eccentric burst prior near merger are not correctly placed.
As the system approaches merger, there is significant GW emission at times other than pericenter crossing, so the isolated bursts assumption breaks as well.

\citetalias{ly2017} did present orbital evolution up to 3rd post-Newtonian (pN) order.
With this proof of concept in place we could improve the eccentric burst prior by extending it to higher pN order.
Improving the centroid mapping equations would allow us to probe bursts closer to the merger and better constrain the astrophysically interesting meta-parameters,  $M$, $Mc$, and $\de_\star$.

Also, by extending to higher pN order the symmetry of the centriod mapping equation is improved.
When truncated to low order a step forward in the evolution equations is not the perfect inverse of a step back.
This is most obvious by examining \autoref{eq:tfor} and \autoref{eq:tback}.
The time step between neighboring bursts appears identical, but when moving from the $i$ to $i+1$ burst and back, the right-hand-sides are evaluated at different bursts in the orbital sequence.
Going to higher order would reduce this problem and potentially lead to better physical parameter recovery.

These two improvements would not address the fact that bursts become less isolated near merger.
To account for this a change of wavelet basis in the signal model could be implemented.
Using a chirplet basis has been shown to improve burst searches for quasi-circular BBHs \cite{2018PhRvD..97j4057M}.
A wavelet could be constructed that better represents the high amplitude, high frequency pericenter emission paired with low amplitude, low frequency medial emission.
A recent analytic effective flyby waveform has been specifically designed to approximate the GW burst emission from highly eccentric binaries \citet{Loutrel:2019kky}.
Additionally, highly eccentric pericenter emission should be well approximated by the emission from a hyperbolic passage, so the GW model of \cite{1977ApJ...216..610T} may provided a useful basis function.

Our method could potentially be applied to data from other GW detectors beyond the current terrestrial interferometers.
As binaries are expected to circularize over time \cite{Peters:1964zz}, accessing lower frequencies and therefore earlier evolution could be a boon for the detection of highly eccentric systems.
Third generation terrestrial detectors, which can probe frequencies a factor of a few to 10 lower than the current, second generation detectors, are expected to detect many more systems with non-negligible eccentricity \cite{Lower:2018seu}.
Similarly, eccentric sources are expected to be an important source for LISA with a fraction of the population entering the LISA band with very high eccentricity, $\sim 1$ \cite{ps2010,sd2018,ds2018}.
The massive black holes that LISA observes will enter the band in some cases years before merger \cite{LISA}.
Because our method is a prior that can be applied to any GW burst search, it could be adapted to work with data from any GW detector suitable for a burst search.

This analysis represents a proof of concept for the use of a physically motivated prior to search for highly eccentric BBH systems.
We did not perform a comprehensive study for a wide range of simulated sources. As stated in \autoref{sec:recov}, we generate a small number of sources using different $M$s.
We did not exhaustively explore the parameter space of masses, mass ratios, or eccentricities.
A future analysis could consider a more realistic population of sources.
This analysis does not use a formal detection statistic.
Instead we opt to use a parameter estimation figure of merit as a simple proxy for detection.
We choose not to compute the full Bayesian evidence because of the computational expense.
A future analysis where our prior is implemented in \code{BayesWave} would be able to compute Bayesian evidence ratios for cases with and without the eccentric burst prior.

We believe that using physically motivated priors is a promising method to target GW burst searches towards particular sources.
The eccentric burst prior could be implemented as part of a follow-up stage in existing GW burst pipelines.
When the pipeline finds a candidate event, this targeted method would search for nearby GW bursts of marginal significance that could have arisen from a highly eccentric binary.
Individually these isolated bursts might not stand above the background, but when taken together their significance grows. 
The code for this work can be found at~\cite{ecc-prior}.

\begin{acknowledgments}
  B.D.C.~would like to thank the West Virginia University Department of Graduate Education and Life for recognition as a STEM Mountains of Excellence Fellow.
This fellowship is funded by a grant provided by the West Virginia Higher Education Policy Commission’s Division of Science and Research.
B.D.C.~would also like to thank the Chancellor's Scholar Program at WVU and the CIFAR Azrieli Global Scholarship in Gravity \& the Extreme Universe.
P.T.B.~was partially supported by the West Virginia University Center for Gravitational Waves and Cosmology.
Both B.D.C.~and P.T.B.~received partial support from National Science Foundation (NSF) Physics Frontier Center award \#1430284.
\par
\end{acknowledgments}

\appendix*
 \section{Components of \texorpdfstring{$J_\pm$}{jacobian}}
\label{app:jacobian}
In order to calculate the Jacobian matrices, $J_+$ and $J_-$, of \autoref{eq:sigma}, we started with the low order series approximations of the centroid mapping equations, \autoref{eq:tffor} and \autoref{eq:tfback}.
The two small parameters of the perturbation series are $\de = 1-e \ll 1$ and $(\pi f \mathcal{M})^{5/3} \ll 1$.
We analytically computed the derivatives of the centroid mapping equations and truncated the resulting series.
Each $J_{\pm}$ matrix was computed as the full $3\times 3$ matrix which included the effects of $\de$ on the centroid uncertainty.
The nine components of both $J_+$ and $J_-$ are presented below.

\begin{widetext}
\begin{subequations}
\begin{align}
    J_{+,tt} &= \frac{\partial t_{i+1}}{\partial t} = 1 \\
    J_{+,tf} &= \frac{\partial t_{i+1}}{\partial f} =
        -\frac{1}{f_i^2}\sqrt{\frac{2-\de_i}{\de_i^3}} \\
    J_{+,t\de} &= \frac{\partial t_{i+1}}{\partial \de} =
        \frac{\de_i-3}{f_i\sqrt{(2 - \de_i)\de_i^5}} \\
    \\
    J_{+,ft} &= \frac{\partial f_{i+1}}{\partial t} = 0 \\
    J_{+,ff} &= \frac{\partial f_{i+1}}{\partial f} =
        1 + \frac{(4140 + 7547 \de_i)\pi}{540\, 2^{2/3}}(\pi f_i \Mc)^{5/3}
        + \frac{3725 \pi^2}{144\, 2^{1/3}}(\pi f_i \Mc)^{10/3}\\
    J_{+,f\de} &= \frac{\partial f_{i+1}}{\partial \de} =
        \frac{7547 \pi f_i}{540\,2^{2/3}}(\pi f_i \Mc)^{5/3}\\
    \\
    J_{+,\de t} &= \frac{\partial \de_{i+1}}{\partial t} = 0 \\
    J_{+,\de f} &= \frac{\partial \de_{i+1}}{\partial f} =
        \frac{17(225 + 121\de_i) \Mc \pi^2}{81\, 2^{2/3}} (\pi f_i \Mc)^{2/3} \\
    J_{+,\de \de} &= \frac{\partial \de_{i+1}}{\partial \de} =
        1 + \frac{2057\pi}{135\, 2^{2/3}} (\pi f_i \Mc)^{5/3}.
\end{align}
\end{subequations}

\begin{subequations}
\begin{align}
    J_{-,tt} &= \frac{\partial t_{i-1}}{\partial t} = 1 \\
    J_{-,tf} &= \frac{\partial t_{i-1}}{\partial f} =
        \frac{1}{f_i^2}\sqrt{\frac{2-\de_i}{\de_i^3}} \\
    J_{-,t\de} &= \frac{\partial t_{i-1}}{\partial \de} = 
        \frac{3-\de_i}{f_i\sqrt{(2 - \de_i)\de_i^5}} \\
    \\
    J_{-,ft} &= \frac{\partial f_{i-1}}{\partial t} = 0 \\
    J_{-,ff} &= \frac{\partial f_{i-1}}{\partial f} = 
        1 - \frac{(4140 + 7547 \de_i)\pi}{405}(\pi f_i \Mc)^{5/3} 
        + \frac{3725 \pi^2}{144\, 2^{1/3}}(\pi f_i \Mc)^{10/3} \\
    J_{-,f\de} &= \frac{\partial f_{i-1}}{\partial \de} = 
        - \frac{7547 \pi f_i}{540\,2^{2/3}}(\pi f_i \Mc)^{5/3} \\
    \\
    J_{-,\de t} &= \frac{\partial \de_{i-1}}{\partial t} = 0 \\
    J_{-,\de f} &= \frac{\partial \de_{i-1}}{\partial f} =
        - \frac{425 \Mc \pi^2}{9\, 2^{2/3}} (\pi f_i \Mc)^{2/3} \\
    J_{-,\de \de} &= \frac{\partial \de_{i-1}}{\partial \de} =
        1 + \frac{2057\pi}{135\, 2^{2/3}} (\pi f_i \Mc)^{5/3}.
\end{align}
\end{subequations}
\end{widetext}

\bibliography{biblio}

\end{document}